\documentclass[amsmath,superscriptaddress,showpacs,showkeys,twocolumn,prl]{revtex4-1}
\usepackage{amssymb,amsmath}   
\usepackage[dvips]{graphicx}   
\usepackage{verbatim}   
\usepackage{color}      
\usepackage{subfigure}  
\usepackage{hyperref}   
\usepackage{gensymb}
\usepackage{epstopdf}
\usepackage{natbib}
\usepackage{braket}
\usepackage{enumerate}

\begin{document}

\title{Stabilization of Qubit Relaxation Rates by Frequency Modulation}
	
\author{Shlomi Matityahu}\email{shlomo.matytyahu2@kit.edu}
\affiliation{Institut f\"ur Theorie der Kondensierten Materie, Karlsruhe Institute of Technology, 76131 Karlsruhe, Germany}
\author{Alexander Shnirman}
\affiliation{Institut f\"ur Theorie der Kondensierten Materie, Karlsruhe Institute of Technology, 76131 Karlsruhe, Germany}
\affiliation{Institut f\"ur Quantenmaterialien und Technologien, Karlsruhe Institute of Technology, 76021 Karlsruhe, Germany}
\author{Moshe Schechter}
\affiliation{Department of Physics, Ben-Gurion University of the Negev, Beer Sheva 84105, Israel}

\date{\today}
\begin{abstract}
Temporal, spectral, and sample-to-sample fluctuations in coherence properties of qubits form an outstanding challenge for the development of upscaled fault-tolerant quantum computers. A ubiquitous source for these fluctuations in superconducting qubits is a set of atomic-scale defects with a two-level structure. Here we propose a way to mitigate these fluctuations and stabilize the qubit performance. We show that frequency modulation of a qubit or, alternatively, of the two-level defects, leads to averaging of the qubit relaxation rate over a wide interval of frequencies.
\end{abstract}
	
\keywords{} \maketitle

{\it Introduction and main results.}
A solid-state quantum computer based on a few tens of superconducting qubits has recently been shown to outperform classical supercomputers~\cite{AF19}, and to perform quantum simulations and computations~\cite{BH17,ZJ17} in the so-called noisy intermediate-scale regime. Increasing the circuit depth (i.e., the number of consecutive gates in an algorithm) and scaling-up these processors requires not only better coherence, but is also a matter of reducing fluctuations in qubit properties, such as transition frequency and relaxation rate~\cite{KPV18,SS19,BJ19}. 

The simplest picture of these fluctuations is based on a model~\cite{MC15,MC19} of 
relatively sparse two-level systems (TLSs), which are near resonance with the qubit~\cite{SRW04,CKB04,MJM05}. If $\Delta$, the average spacing between the transition energies of TLSs, is larger than 
$\gamma^{}_{\mathrm{typ}}$, the typical width of the TLS absorption due to phonons, the relaxation rate of the qubit as a function of energy is a collection of randomly distributed Lorentzians, acquiring thus a quasi-random pattern. Moreover, due to spectral diffusion of near-resonant TLSs, these Lorentzians shift on long timescales, giving rise to temporal fluctuations of the qubit relaxation rate. The spectral diffusion is a consequence of the interaction between high-frequency (near-resonant) and low-frequency TLSs that fluctuate between their states due to thermal activation~\cite{MC15,FL15,MS18}. Due to the wide range of TLS relaxation times~\cite{PWA72,AHV72}, the fluctuations in qubit properties range from milliseconds to hours and days.

In this letter, we consider a qubit coupled to a set of TLSs and subject to a periodic (harmonic) modulation of its frequency, i.e., a variant of the Tien-Gordon setup~\cite{TG63}. Each TLS can spontaneously relax from its excited state by emitting a phonon. The Hamiltonian of the system reads $\mathcal{H}^{}_{}(t)=\mathcal{H}^{}_{\mathrm{Qubit-TLS}}(t)+\mathcal{H}^{}_{\mathrm{TLS-Phonons}}+\mathcal{H}^{}_{\mathrm{Phonons}}$, where $\mathcal{H}^{}_{\rm TLS-Phonons}$ describes the coupling of the TLSs to phonons and
\begin{widetext}
\begin{align}
\label{eq:Hamiltonian}
\mathcal{H}^{}_{\mathrm{Qubit-TLS}}(t)=&-\frac{1}{2}\left[E^{}_{0}+A\cos(\Omega t)\right]\sigma^{}_{z}-\frac{1}{2}\sum^{}_{n}\varepsilon^{}_{n}\tau^{(n)}_{z}+\frac{1}{2}\sum^{}_{n}g^{}_{n}\sigma^{}_{x}\tau^{(n)}_{x}.
\end{align}
\end{widetext}
Here, $\sigma^{}_{x}$, $\sigma{}_{z}$ and $\tau^{(n)}_{x}$, $\tau^{(n)}_{z}$ are the Pauli matrices describing the qubit and the $n$-th TLS. The first term describes a qubit with energy splitting $E^{}_{0}$, modulated harmonically with frequency $\Omega$ and amplitude $A$, and the second term describes a set of TLSs with energy splittings $\varepsilon^{}_{n}$. The third term is a transverse coupling between the qubit and the TLSs with strengths $g^{}_{n}$. Each TLS is characterized by the relaxation rate $\gamma_n$ due to the coupling to phonons, as will be discussed below.

Similarly to Ref.~\cite{TG63}, we show that the periodic driving splits the TLS absorption peaks into multiple replications shifted by multiples of the driving frequency $\Omega$. The total weight of the replicated peaks is equal to that of the original peak. Thus, assuming a constant spectral density of TLSs on a large energy scale, the average relaxation rate of the qubit is not modified. Yet, the replication effectively reduces the spacing between the peaks, thus reducing fluctuations of the qubit relaxation rate. We find the spread of qubit relaxation rates to be reduced by a factor $\sim \sqrt{\Omega/A}$ for $A>\Omega$.

A similar problem was analyzed by Agarwal~\cite{AGS99}, who considered the modulation of the coupling of an atom to a heat bath. In terms of the present work, the results of Ref.~\cite{AGS99} correspond to a single weakly coupled TLS characterized by an energy splitting $\varepsilon$ and $g\ll\gamma$, at resonance with the qubit ($E^{}_{0}=\varepsilon$). A periodic driving with $A>\Omega>\gamma$ splits the absorption peak of the TLS into multiple weaker ones, so that the relaxation of the qubit is slowed down.

Here we study the relaxation of a qubit coupled to a bath of spectrally sparse TLSs, i.e., $\Delta>\gamma^{}_{n},g^{}_{n}$. We consider both the weak-coupling ($g^{}_{n}\ll\gamma^{}_{n}$) and strong-coupling ($g^{}_{n}\gg\gamma^{}_{n}$) regimes. The parameter regime $A\gg\Delta>\gamma^{}_{n}, g^{}_{n}$ is assumed, such that the qubit crosses resonances with multiple TLSs, which leads to the averaging effect. We also assume $\Omega\gg\gamma^{}_{n}, g^{}_{n}$, meaning that multiple resonances with each TLS are crossed coherently (alternatively, this condition means that the replications of each absorption peak extend beyond its width). In the weak-coupling regime, we find an analytical expression for the qubit relaxation rate as a function of its central energy splitting $E^{}_{0}$, and analyze its statistics. We show that the average rate remains the same as in the undriven case, but fluctuations of it are reduced. In the strong-coupling regime we show numerically that frequency modulation with $A/\Omega\gtrsim(g/\gamma)^{2}$ effectively transforms a strongly coupled TLS into a series of $~2A/\Omega$ weakly coupled TLSs separated by energy $\Omega$. Fluctuations are reduced also in this case.

{\it Derivation and results.}
By integrating out the phonon degrees of freedom in a Weisskopf-Wigner theory~\cite{WV30,Supp}, we find that at $T=0$ the exact description of the qubit relaxation dynamics is given by the following non-hermitian Hamiltonian~\cite{Remark1}:
\begin{align}
\label{eq:NonHermHamiltonian}
\mathcal{H}^{}_{nh}(t)=\mathcal{H}^{}_{\mathrm{Qubit-TLS}}(t)-\frac{i}{2}\sum^{}_{n}\gamma^{}_{n}(1-\tau^{(n)}_{z}).
\end{align}
The non-Hermitian term describes TLSs decaying from their excited states (relaxation rates $\gamma^{}_{n}$ for the amplitude or $2\gamma^{}_{n}$ for the probability). The initial state of the system is $\ket{\psi(0)}=\ket{1}\ket{0,\ldots,0}$, in which the qubit is excited and the TLSs are at their ground states. 

Let us briefly discuss the static case ($A=0$) for the Hamiltonian~(\ref{eq:NonHermHamiltonian}). For a single TLS, one can diagonalize the Hamiltonian in the subspace $\{\ket{0}\ket{1},\ket{1}\ket{0}\}$ to obtain the two eigenvalues $E^{}_{\pm}=\pm\sqrt{(E^{}_{0}-\varepsilon+i\gamma)^{2}+g^{2}}/2-i\gamma/2$. In the limit $g\ll |E^{}_{0}-\varepsilon|$ or $g\ll\gamma$, the coupling is weak and the qubit acquires a finite decay rate (of the probability to be in the excited state) given by the imaginary part of $E^{}_{+}$,
\begin{align}
\label{eq:Gamma_static}
\Gamma^{}_{\mathrm{static}}(E^{}_{0})=-2\operatorname{Im}(E^{}_{+})=\frac{1}{2}\frac{g^{2}\gamma}{\left(E^{}_{0}-\varepsilon\right)^{2}+\gamma^{2}}.
\end{align}
In the strong-coupling limit, $g\gg |E^{}_{0}-\varepsilon|, \gamma$, the system oscillates between the states $\ket{1}\ket{0}$ and $\ket{0}\ket{1}$ and decays with rate $\gamma$.

To analyze the effect of a periodic frequency modulation we employ again the 
Weisskopf-Wigner approach~\cite{WV30}. We treat the last term of the Hamiltonian~(\ref{eq:Hamiltonian}) as a perturbation, use the rotating wave approximation, and consider the general single excitation state $\ket{\psi^{}_{\mathrm{I}}(t)}=a(t)\ket{1}\ket{0,\ldots,0}+\sum^{}_{n}b^{}_{n}(t)\ket{0}\ket{0,\ldots,0,1^{}_{n},0,\ldots,0}$ in the interaction picture. The corresponding Schr\"{o}dinger equation reads
\begin{align}
\label{eq:Schroedinger}
&\dot{a}(t)=-\frac{i}{2}\sum^{}_{n}g^{}_{n}e^{i\left[E^{}_{0}t+\phi(t)-(\varepsilon^{}_{n}-i\gamma^{}_{n})t\right]}b^{}_{n}(t),\nonumber\\
&\dot{b}^{}_{n}(t)=-\frac{i}{2}g^{}_{n}e^{-i\left[E^{}_{0}t+\phi(t)-(\varepsilon^{}_{n}-i\gamma^{}_{n})t\right]}a(t),
\end{align}  
where $\phi(t)=\int^{t}_{0}A\cos(\Omega t')dt'=\left(A/\Omega\right)\sin(\Omega t)$. Note that we are interested in the regime $\Omega>\gamma^{}_{n}$, such that the qubit experiences multiple resonant passages with the TLS on the time scale $1/\gamma^{}_{n}$. Using the Fourier series expansion $e^{i\phi(t)}=\sum^{\infty}_{m=-\infty}J^{}_{m}(A/\Omega)e^{im\Omega t}$, where $J^{}_{m}(x)$ is the Bessel function of the first kind, we obtain the following integro-differential equation:
\begin{widetext}
\begin{align}
\label{eq:a(t)1}
\dot{a}(t)=&-\frac{1}{4}\sum^{}_{n}\sum^{\infty}_{m^{}_{1},m^{}_{2}=-\infty}g^{2}_{n}J^{}_{m^{}_{1}}J^{}_{m^{}_{2}}e^{i(m^{}_{1}-m^{}_{2})\Omega t}\int^{t}_{0}e^{i\left(E^{}_{0}+m^{}_{2}\Omega-\varepsilon^{}_{n}\right)\tau}e^{-\gamma^{}_{n}\tau}a(t-\tau)d\tau,
\end{align}
\end{widetext}
where the Bessel functions are evaluated at $A/\Omega$.

In the weak-coupling limit, $g^{}_{n}\ll\gamma^{}_{n}$, the qubit relaxation rate on resonance with the $n$-th TLS and without frequency modulation ($A=0$) is estimated as $\Gamma^{}_{\mathrm{static}}\approx g^{2}_{n}/\gamma^{}_{n}\ll\gamma^{}_{n}$ [Eq.~(\ref{eq:Gamma_static})]. Since frequency modulation is expected to reduce this rate, we conclude that $a(t)$ varies slowly on the time scale $1/\gamma^{}_{n}$, and we can safely invoke the Markov approximation, i.e.\ we replace $a(t-\tau)$ by $a(t)$ and extend the upper limit of the integration to $\infty$ (note that the justification here for the Markov approximation is different from that in the Weisskopf-Wigner theory for spontaneous emission, where an atom decays into a continuum of radiation modes. Here the TLSs do not form a continuum by themselves, but rather a structured continuum due to their relaxation to phonons. The Markov approximation is justified because the qubit relaxation rate is smaller than that of a single TLS, i.e., $\gamma_n$). Moreover, the condition $\Omega>\gamma^{}_{n}$ implies that $e^{i(m^{}_{1}-m^{}_{2})\Omega t}$ oscillates rapidly unless $m^{}_{1}=m^{}_{2}$. Equation~(\ref{eq:a(t)1}) therefore reduces to $\dot{a}(t)=-Ca(t)$, where
\begin{align}
\label{eq:a(t)2}
C=&\frac{1}{4}\sum^{}_{n}\sum^{\infty}_{m=-\infty}\frac{g^{2}_{n}J^{2}_{m}}{\gamma^{}_{n}-i\left(E^{}_{0}+m\Omega-\varepsilon^{}_{n}\right)}.
\end{align}
The qubit relaxation rate $\Gamma$ is given by the decay rate of $|a(t)|^2$, i.e.\
\begin{align}
\label{eq:Gamma}
\Gamma(E^{}_{0})=2\operatorname{Re}(C)=\frac{1}{2}\sum^{}_{n}\sum^{\infty}_{m=-\infty}\frac{g^{2}_{n}J^{2}_{m}\gamma^{}_{n}}{\left(E^{}_{0}+m\Omega-\varepsilon^{}_{n}\right)^{2}+\gamma^{2}_{n}}.
\end{align}

The Lorentzian contributed by each TLS in the absence of frequency modulation, Eq.~(\ref{eq:Gamma_static}), is thus replicated at energy intervals of $\Omega$. The width of each replication is again $\gamma^{}_{n}$, but its height is reduced and given by $\tilde{g}^{2}_{nm}/\gamma^{}_{n}$, with $\tilde{g}^{}_{nm}=g^{}_{n}J^{}_{m}(A/\Omega)<g^{}_{n}$. Since $J^{}_{m}(x\rightarrow 0)\rightarrow\delta^{}_{m,0}$, Eq.~(\ref{eq:Gamma}) reduces to a sum of Lorentzians (one per each TLS) of the form~(\ref{eq:Gamma_static}) in the limit $A/\Omega\ll 1$; the effect of frequency modulation is therefore non-trivial only in the limit $A/\Omega\gg 1$. We also note that for a given $x$, $J^{}_{m}(x)$ approaches zero for $|m|\gtrsim |x|$, and thus the number of replications for each TLS is $\approx 2A/\Omega$.  

To compare the qubit relaxation rate with and without frequency modulation, we now calculate its average value over different realizations of the TLSs, as well as its variance. Consider $N$ TLSs with random independent energy splittings $\varepsilon^{}_{n}$ homogeneously distributed in the energy range $\left[-E/2,E/2\right]$, assuming the thermodynamic limit $N, E\rightarrow\infty$ with a finite average level spacing $\Delta=E/N$ between TLSs. The joint probability distribution function is $P(\varepsilon^{}_{1},\ldots,\varepsilon^{}_{N})=1/E^{N}$ and the first moment (average) of $\Gamma$ [Eq.~(\ref{eq:Gamma})] is given, thus, by
\begin{align}
\label{eq:Gamma_av}
\braket{\Gamma}&=\int^{E/2}_{-E/2}d\varepsilon^{}_{1}\ldots\int^{E/2}_{-E/2}d\varepsilon^{}_{N}P(\varepsilon^{}_{1},\ldots,\varepsilon^{}_{N})\Gamma(\varepsilon^{}_{1},\ldots,\varepsilon^{}_{N})\nonumber\\
&=\frac{1}{2E}\sum^{N}_{n=1}\sum^{\infty}_{m=-\infty}\int^{E/2}_{-E/2}\frac{g^{2}_{n}J^{2}_{m}\gamma^{}_{n}}{\left(E^{}_{0}+m\Omega-\varepsilon^{}_{n}\right)^{2}+\gamma^{2}_{n}}d\varepsilon^{}_{n}\nonumber\\
&=\frac{\pi\braket{g^{2}}}{2\Delta},
\end{align}
where $\braket{g^{2}}=(1/N)\sum^{N}_{n=1}g^{2}_{n}$. In the last step we took the the limit $E\rightarrow\infty$ and used the relation $\sum^{\infty}_{m=-\infty}J^{2}_{m}(x)=1$. Note that $\braket{\Gamma}$ does not depend on $A$, and therefore is not affected by frequency modulation. Similarly, the variance of $\Gamma$ is given by
\begin{align}
\label{eq:Gamma_sq}
\sigma^2\equiv \braket{\Gamma^{2}}-\braket{\Gamma}^{2}\approx\frac{\pi}{8\Delta}\Big\langle\frac{g^{4}}{\gamma}\Big\rangle\!\sum^{\infty}_{m=-\infty}J^{4}_{m}(A/\Omega),
\end{align}
where $\braket{g^{4}/\gamma}=(1/N)\sum^{N}_{n=1}g^{4}_{n}/\gamma^{}_{n}$. In Eq.~(\ref{eq:Gamma_sq}) we neglected a sub-leading contribution carrying an additional small factor of the order of $\gamma^{2}_{\mathrm{typ}}/\Omega^{2}$, originating in the overlaps of Lorentzians with different values of $m$. In the absence of frequency modulation ($A=0$), one has $J^{}_{m}(0)=\delta^{}_{m,0}$ and Eq.~(\ref{eq:Gamma_sq}) simplifies to $\sigma^2_{\mathrm{static}}=(\pi/8\Delta)\braket{g^{4}/\gamma}$. Replacing $g$ and $\gamma$ by their typical values, we find that the relative standard deviation in the static case is $\sigma^{}_{\mathrm{static}}/\braket{\Gamma}=\sqrt{1/2\pi}\sqrt{\Delta/\gamma^{}_{\mathrm{typ}}}$. This means that fluctuations in $\Gamma$ are large for $\Delta\gg\gamma^{}_{\mathrm{typ}}$, i.e., if the TLSs are sparsely distributed in energy. In the dynamic case with $A/\Omega\gg 1$, the sum in Eq.~(\ref{eq:Gamma_sq}) scales as $\Omega/A$~\cite{Remark2}, and thus the standard deviation decreases as $\sqrt{\Omega/A}$ compared to the static result, i.e.
\begin{align}
\label{eq:std_ratio}
\frac{\sigma^{}_{\mathrm{modulation}}}{\sigma^{}_{\mathrm{static}}}\propto\sqrt{\frac{\Omega}{A}}.
\end{align} 
The meaning of this result is that fluctuations in $\Gamma$ are expected to reduce as $\sqrt{\Omega/A}$ due to the periodic driving of the qubit frequency.

\begin{figure}[ht!]
	\includegraphics[width=0.5\textwidth,height=0.25\textheight]{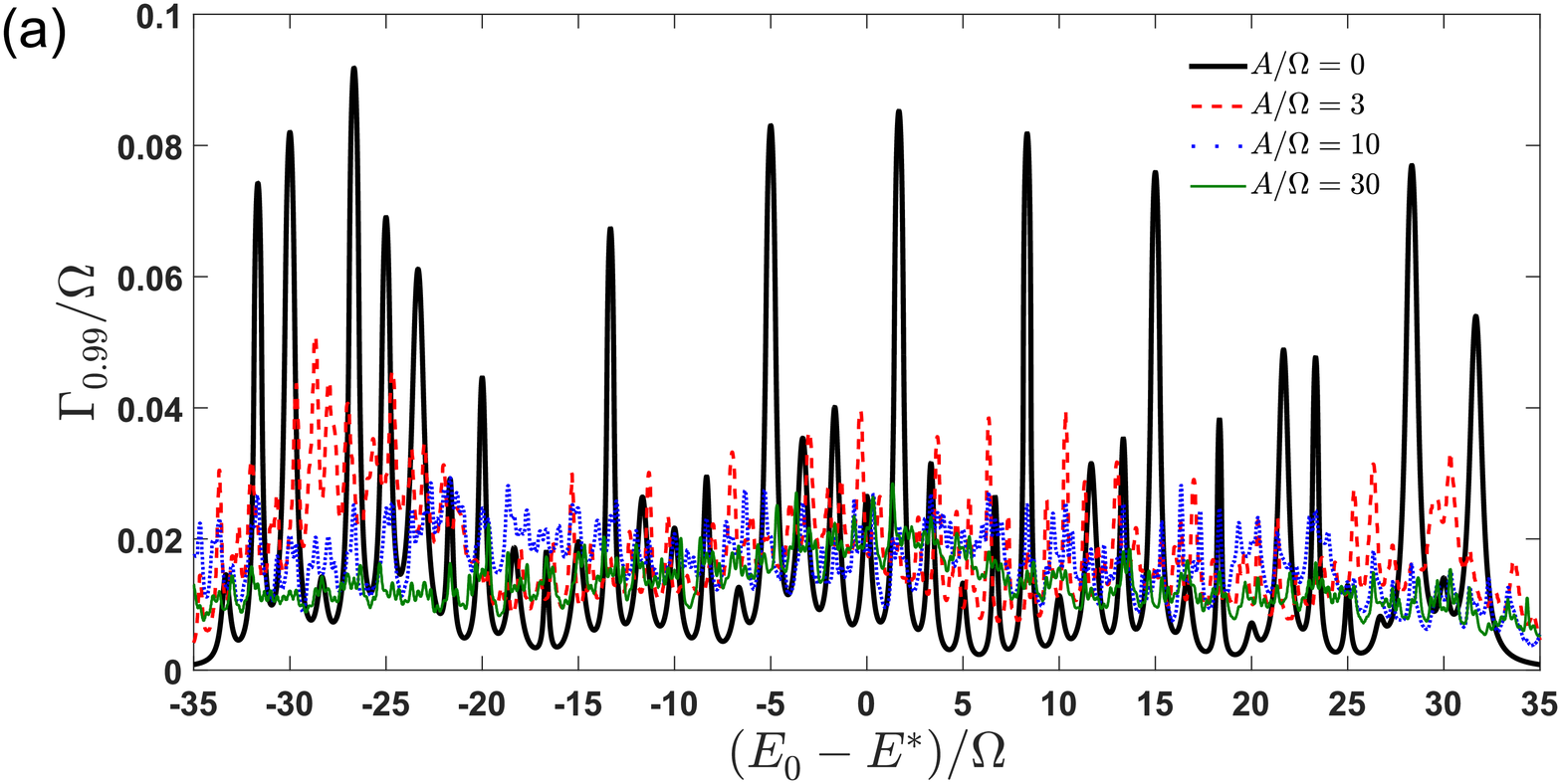}
	\includegraphics[width=0.5\textwidth,height=0.25\textheight]{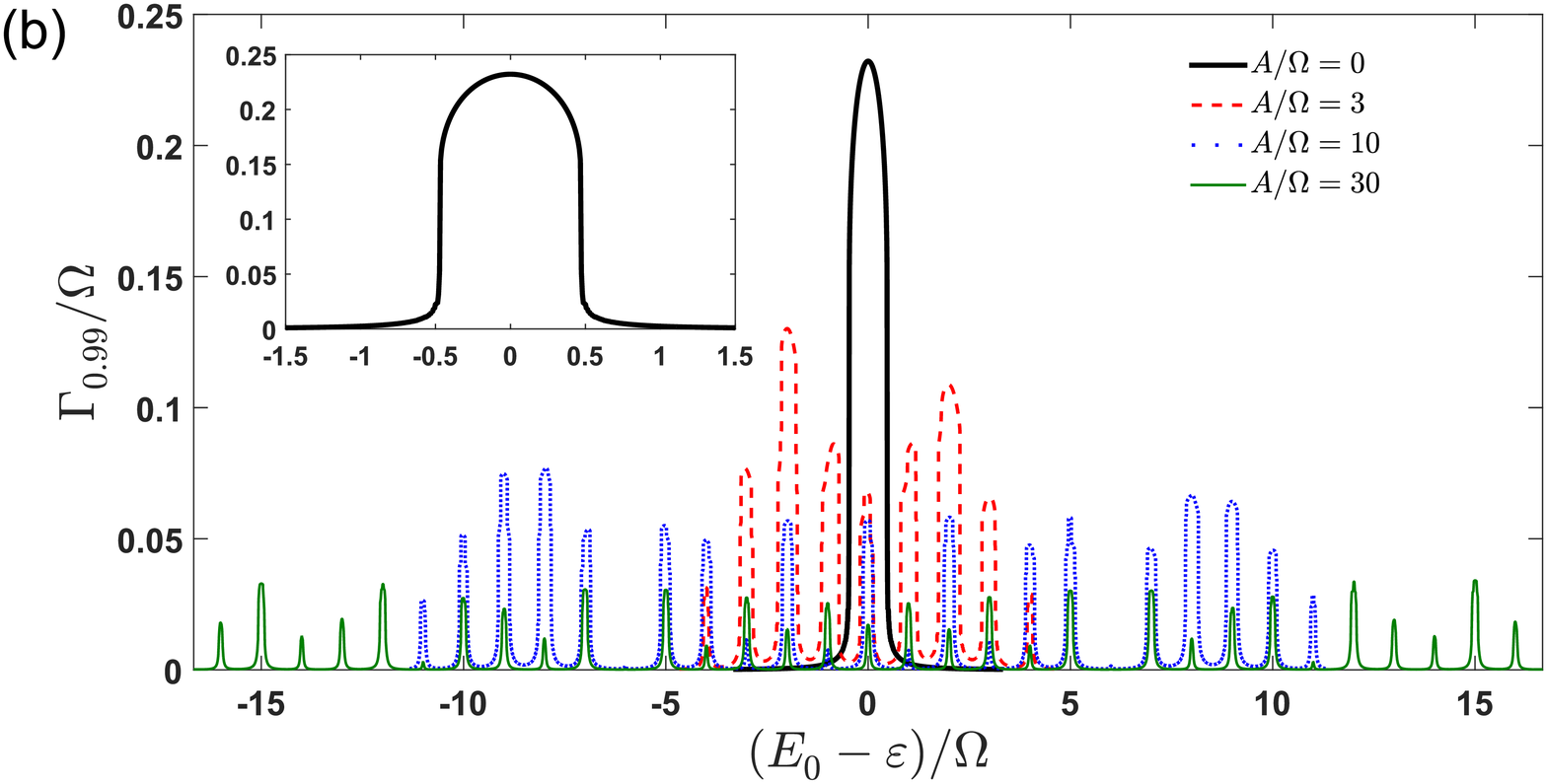}
	\caption{(Color online) Qubit relaxation rate as a function of its central energy splitting $E^{}_{0}$, obtained from a numerical integration of the Schr\"{o}dinger equation~(\ref{eq:Schroedinger}) for various values of $A/\Omega$, as indicated in the legend. The relaxation rate $\Gamma^{}_{0.99}$ (in units of $\Omega$) is calculated as $1/T$, where $T$ is the time for which $|a(T)|=0.99$. (a) For a bath of 40 weakly coupled TLSs with a constant level spacing $\Delta/\Omega=5/3$. The couplings $g^{}_{n}/\Omega$ and relaxation rates $\gamma^{}_{n}/\Omega$ are randomly chosen from homogeneous distributions in the ranges $[2/3,10/3]\cdot 10^{-2}$ and $[2/3,10/3]\cdot 10^{-1}$, respectively. The qubit central energy splitting is counted relative to some typical energy splitting of the qubit, $E^{\ast}$. (b) For a single strongly coupled TLS, with $g/\Omega=2/3\cdot 10^{-1}$ and $\gamma/\Omega=0.02$. The inset is a zoom-in of the undriven case ($A=0$), showing a flat peak with height $\propto g$ for $|E^{}_{0}-\varepsilon|<g$ and Lorentzian tails for $|E^{}_{0}-\varepsilon|>g$ (note that $\varepsilon$ is taken as the origin of the horizontal axis).}
		\label{fig1}
\end{figure}

In Fig.~\ref{fig1}(a) we plot the result of a numerical integration of the Schr\"{o}dinger equation~(\ref{eq:Schroedinger}) for 40 weakly coupled TLSs ($g^{}_{n}\ll\gamma^{}_{n}$) with equispaced energy splittings. The couplings and relaxation rates are randomly selected from homogeneous distributions. The qubit relaxation rate is calculated as $\Gamma^{}_{0.99}\equiv 1/T$, where $T$ is the time for which $|a(T)|=0.99$. Due to frequency modulation, the peaks corresponding to the undriven case ($A=0$) are replicated at multiples of $\Omega$, resulting in diminishing fluctuations in the driven case.

Let us discuss now the strong-coupling regime $g^{}_{n}\gg\gamma^{}_{n}$. The Markov approximation in Eq.~(\ref{eq:a(t)1}) is not applicable in this case, but one can still restrict the sum over $m^{}_{1}$ and $m^{}_{2}$ to $m^{}_{1}=m^{}_{2}$, provided that $\Omega$ is much larger than the typical coupling. The effect of frequency modulation is again to replicate the peak of each TLS at multiples of $\Omega$, with the coupling corresponding to the replication at $\varepsilon^{}_{n}+m\Omega$ being $\tilde{g}^{}_{nm}=g^{}_{n}J^{}_{m}(A/\Omega)<g^{}_{n}$. This effectively transforms the contribution of a single strongly coupled TLS with energy splitting $\varepsilon^{}_{n}$ to a collection of TLSs at energy splittings $\tilde{\varepsilon}^{}_{nm}=\varepsilon^{}_{n}+m\Omega$ with smaller couplings $\tilde{g}^{}_{nm}$. Since $J^{}_{m}(A/\Omega)\sim\sqrt{\Omega/A}$ for $|m|<A/\Omega$, all the replications correspond to the weak coupling regime for $A/\Omega\gtrsim(g/\gamma)^{2}$. This is demonstrated in Fig.~\ref{fig1}(b) showing the result of a numerical integration of the Schr\"{o}dinger equation~(\ref{eq:Schroedinger}) for a single strongly coupled TLS. In the undriven case, one observes the two different behaviors corresponding to $|E^{}_{0}-\varepsilon|<g$ and $|E^{}_{0}-\varepsilon|>g$. For $|E^{}_{0}-\varepsilon|<g$, the qubit is in resonance with the TLS and the system oscillates between the states $\ket{1}\ket{0}$ and $\ket{0}\ket{1}$ with frequency $g$, while these oscillations decay with rate $\gamma$. For quantum computing purposes, we must identify the effective relaxation rate in this case as $\sim g$ rather then $\gamma$. Indeed, unless very specific for each TLS protocols are employed, the qubit state is lost already during the first oscillation period. On the other hand, outside of resonance ($|E^{}_{0}-\varepsilon|>g$) the coupling is weak and the decay of the qubit population is again exponential with rate given by Eq.~(\ref{eq:Gamma_static}). Such a flat peak with height and width equal to $g$, and with Lorentzian tails for $|E^{}_{0}-\varepsilon|>g$, is seen in the undriven case corresponding to the black (bold solid) line in Fig.~\ref{fig1}(b). This peak is replicated at multiples of $\Omega$ in the presence of frequency modulation, and for $A/\Omega\gtrsim(g/\gamma)^{2}=9$ the replications are described by the Lorentzians corresponding to the weak-coupling regime. If $\Omega$ is comparable to the typical coupling, one cannot neglect terms with $m^{}_{1}\neq m^{}_{2}$ in Eq.~(\ref{eq:a(t)1}). The replication at $\varepsilon^{}_{n}+m\Omega$ then consists of several contributions of $m^{}_{1}$, $m^{}_{2}$ satisfying $m^{}_{1}-m^{}_{2}=m$. This leads to more complex pattern of replicated peaks, see an example in the supplemental material~\cite{Supp}.

{\it Discussion.}
Equivalently, and even more beneficially for the purposes of quantum computation, one could modulate the frequencies of the TLSs. As shown in several works, the energy splitting of TLSs can be manipulated by external strain and electric fields~\cite{GGJ12,KMS14,LJ15,LJ16,SB16,BJD17,LJ19,BA20,BA21}. In a recent paper~\cite{MS19}, we have studied the dielectric loss of a superconducting micro-resonator in the presence of a periodic frequency modulation of near-resonant TLSs. If the modulation frequency $\Omega$ is smaller than the TLS relaxation rate $\gamma_n\sim 1\,$MHz, resonant passages of each TLS are independent and the loss increases with $\Omega$ up to its intrinsic low-power value~\cite{KMS14,BAL13}. In the regime $\Omega>\gamma^{}_{n}$, consecutive resonant passages form a coherent series of Landau-Zener transitions. The interference pattern shows narrow absorption peaks (due to constructive interference) and wide domains of reduced absorption (due to destructive interference) as a function of the relative phase between transitions. Upon averaging over the ensemble of TLSs, the destructive interference dominates as $\Omega$ increases, which leads to decrease of the resonator loss~\cite{MS19}. The replicated peaks in the present letter correspond exactly to the constructive interference peaks of Ref.~\cite{MS19}. Whereas the physics discussed in Ref.~\cite{MS19} is dominated by TLS saturation at high powers (i.e., resonators at the classical regime), the results in the present letter concern the opposite limit of a single excitation, relevant to qubits and resonators in the single-photon regime.

Modulating the frequencies of the TLSs would allow performing 1-qubit gates and coupling/decoupling qubits (2-qubit gates) as usual. However, if modulation of the frequency of the qubits is unavoidable, and such modulations should continue during qubit gates, it is necessary to adjust the pulses of the gates accordingly. Consider for example the 1-qubit gates performed by driving resonantly (driving frequency $\omega^{}_{\mathrm{d}}=E^{}_{0}$) Rabi oscillations of the qubit. It is obvious that the effective Rabi frequency would be given by $\Omega^{}_{\mathrm{R},0}=\Omega^{}_{\mathrm{R}}J^{}_{0}(A/\Omega)$, where $\Omega^{}_{\mathrm{R}}$ is the Rabi frequency at $A=0$ (see further details in the supplemental material~\cite{Supp}). One can even induce Rabi oscillations by driving at $\omega^{}_{\mathrm{d}}=E^{}_{0}+m\Omega$ with the effective Rabi frequency given by $\Omega^{}_{\mathrm{R},m}=\Omega^{}_{\mathrm{R}}J^{}_{m}(A/\Omega)$. Thus, at a given $A/\Omega$ one can optimize the Rabi frequency by choosing $m$ with the largest value of $J^{}_{m}$. Inevitably, the effective Rabi frequency is reduced, and hence 1-qubit gates become longer or more power has to be applied. Alternatively, one can apply a Rabi-pulse with multiple harmonics $E^{}_{0}+m\Omega$ and phases chosen so that the effective Rabi amplitudes interfere constructively. Similar considerations apply for protocols coupling the qubits for performing 2-qubit gates.

In addition to relaxation, the TLSs generate an energy shift of the qubit (Lamb shift). This can be estimated from Eq.~(\ref{eq:a(t)2}) as $\Delta E=\operatorname{Im}(C)$. Assuming a homogeneous density of states of the TLSs, one can roughly distinguish two contributions: $\Delta E=\Delta E^{}_{\mathrm{low}}+\Delta E^{}_{\mathrm{high}}$. 
The low frequency part, $\Delta E^{}_{\mathrm{low}}$, is given by TLSs with $0<\varepsilon^{}_{n}<2E^{}_{0}$. This part vanishes upon averaging over realizations of the TLSs. The high frequency part, $\Delta E^{}_{\mathrm{high}}$, is the usual logarithmically diverging Lamb shift, which does not fluctuate much. We therefore include $\Delta E^{}_{\mathrm{high}}$ into $E^{}_{0}$ and focus on $\Delta E^{}_{\mathrm{low}}$. A simple calculation gives $\braket{\Delta E^{}_{\mathrm{low}}}=0$ and 
$\braket{(\Delta E^{}_{\mathrm{low}})^2}\approx (1/4)\mathrm{Var}(\Gamma)$ [cf.\ Eq.~(\ref{eq:Gamma_sq})]. Frequency modulations thus reduce $\braket{(\Delta E^{}_{\mathrm{low}})^2}$, and hence the pure dephasing of the qubit generated by the spectral diffusion of the TLSs.

S.M. acknowledges support by the A. von Humboldt foundation, A.S. acknowledges support by the Baden-W\"urttemberg Stiftung (Project "QuMaS"), and M.S. acknowledges support by the Israel Science Foundation (Grant No. 2300/19).

\newpage
\onecolumngrid
\newcommand{\beginsupplement}{%
	\setcounter{equation}{0}
	\renewcommand{\theequation}{S\arabic{equation}}%
	\setcounter{table}{0}
	\renewcommand{\thetable}{S\arabic{table}}%
	\setcounter{figure}{0}
	\renewcommand{\thefigure}{S\arabic{figure}}%
}
	\beginsupplement	
	\title{Supplemental material for \\ Stabilization of Qubit Relaxation Rates by Frequency Modulation}
	
	\author{Shlomi Matityahu}
	\affiliation{Institut f\"ur Theorie der Kondensierten Materie, Karlsruhe Institute of Technology, 76131 Karlsruhe, Germany}
	\author{Alexander Shnirman}
	\affiliation{Institut f\"ur Theorie der Kondensierten Materie, Karlsruhe Institute of Technology, 76131 Karlsruhe, Germany}
	\affiliation{Institut f\"ur Quantenmaterialien und Technologien, Karlsruhe Institute of Technology, 76021 Karlsruhe, Germany}
	\author{Moshe Schechter}
	\affiliation{Department of Physics, Ben-Gurion University of the Negev, Beer Sheva 84105, Israel}
	
	\date{\today}
	
\section{Supplemental Material}	
	\subsection{Derivation of the non-Hermitian Hamiltonian} \label{Model} The full Hamiltonian of our system reads $\mathcal{H}^{}_{}(t)=\mathcal{H}^{}_{\mathrm{Qubit-TLS}}(t)+\mathcal{H}^{}_{\mathrm{TLS-Phonons}}+\mathcal{H}^{}_{\mathrm{Phonons}}$, where $\mathcal{H}^{}_{\mathrm{Qubit-TLS}}(t)$ is given by Eq.~(1) of the main text and
	\begin{align}
	\label{eq:1}&\mathcal{H}^{}_{\mathrm{TLS-Phonons}}=\frac{1}{2}\sum^{}_{n,k}\tau^{(n)}_{x}(v^{}_{nk}a^{}_{k}+v^{\ast}_{nk}a^{\dag}_{k})\ ,\nonumber\\
	&\mathcal{H}^{}_{\mathrm{Phonons}}=\sum^{}_{k}\omega^{}_{k}a^{\dag}_{k}a^{}_{k}\ .
	\end{align}
	Here $a^{\dag}_{k}$ and $a^{}_{k}$ are the phonon creation and annihilation operators, $\omega^{}_{k}$ are the phonon frequencies, and $v^{}_{nk}$ describes the coupling between the $n$th TLS and the phonon mode with wave vector $k$. We now consider the interaction picture with respect to the perturbation $V=(1/2)\sum^{}_{n}g^{}_{n}\sigma^{}_{x}\tau^{(n)}_{x}+(1/2)\sum^{}_{n,k}v^{}_{nk}\tau^{(n)}_{x}(a^{}_{k}+a^{\dag}_{k})$ and employ the rotating wave approximation. Within the single excitation subspace spanned by the states $\ket{1}\ket{\{0\}^{}_{\mathrm{TLS}}}\ket{\{0\}^{}_{\mathrm{Phonons}}}$, $\ket{0}\ket{\{1_{n}\}^{}_{\mathrm{TLS}}}\ket{\{0\}^{}_{\mathrm{Phonons}}}$ and $\ket{0}\ket{\{0\}^{}_{\mathrm{TLS}}}\ket{\{1_{k}\}^{}_{\mathrm{Phonons}}}$, in which either the qubit, a single TLS, or a single phonon is excited (we use the notation $\ket{\{1_{n}\}^{}_{\mathrm{TLS}}}=\ket{0,\ldots,0,1_{n},0,\ldots,0}$ for the state in which the $n$th TLS is excited and all other TLSs are at their ground states and $\ket{\{1_{k}\}^{}_{\mathrm{Phonons}}}=\ket{0,\ldots,0,1_{k},0,\ldots,0}$ for the state with a single phonon occupying the mode with wave vector $k$). The Schr\"{o}dinger equation for the general state $\ket{\psi^{}_{\mathrm{I}}(t)}=a(t)\ket{1}\ket{\{0\}^{}_{\mathrm{TLS}}}\ket{\{0\}^{}_{\mathrm{Phonons}}}+\sum^{}_{n}b^{}_{n}(t)\ket{0}\ket{\{1_{n}\}^{}_{\mathrm{TLS}}}\ket{\{0\}^{}_{\mathrm{Phonons}}}+\sum^{}_{k}c^{}_{k}(t)\ket{0}\ket{\{0\}^{}_{\mathrm{TLS}}}\ket{\{1_{k}\}^{}_{\mathrm{Phonons}}}$ then reads
	\begin{align}
	\label{eq:2}
	&\dot{a}(t)=-\frac{i}{2}\sum^{}_{n}g^{}_{n}e^{i\left(E^{}_{0}t+\phi(t)-\varepsilon^{}_{n}t\right)}b^{}_{n}(t),\nonumber\\
	&\dot{b}^{}_{n}(t)=-\frac{i}{2}\left[g^{}_{n}e^{-i\left(E^{}_{0}t+\phi(t)-\varepsilon^{}_{n}t\right)}a(t)+\sum^{}_{k}v^{}_{nk}e^{i(\varepsilon^{}_{n}-\omega^{}_{k})t}c^{}_{k}(t)\right],\nonumber\\
	&\dot{c}^{}_{k}(t)=-\frac{i}{2}\sum^{}_{l}v^{\ast}_{lk}e^{-i(\varepsilon^{}_{l}-\omega^{}_{k})t}b^{}_{l}(t),
	\end{align}
	with initial conditions $a(0)=1$, $b^{}_{n}(0)=c^{}_{k}(0)=0\;\; \forall\, n,k$. Integrating the last equation and substituting into the second, we obtain
	\begin{align}
	\label{eq:3}
	&\dot{a}(t)=-\frac{i}{2}\sum^{}_{n}g^{}_{n}e^{i\left(E^{}_{0}t+\phi(t)-\varepsilon^{}_{n}t\right)}b^{}_{n}(t),\nonumber\\
	&\dot{b}^{}_{n}(t)=-\frac{i}{2}g^{}_{n}e^{-i\left(E^{}_{0}t+\phi(t)-\varepsilon^{}_{n}t\right)}a(t)-\frac{1}{4}\sum^{}_{l}e^{i(\varepsilon^{}_{n}-\varepsilon^{}_{l})t}\int^{t}_{0}K^{}_{nl}(t-t')b^{}_{l}(t'),
	\end{align}
	where $K^{}_{nl}(\tau)=\sum^{}_{k}v^{}_{nk}v^{\ast}_{lk}e^{i(\varepsilon^{}_{l}-\omega^{}_{k})\tau}$. We now apply the Weisskopf-Wigner (Markov) approximation~\cite{WV30}.
	This is justified because the functions $K^{}_{nl}(\tau)$ decay rapidly on the characteristic time scale $1/\gamma^{}_{l}$ (to be defined below) describing the temporal variation of $b^{}_{l}(t)$. One then replaces $b^{}_{l}(t')$ by $b^{}_{l}(t)$ in the integral and extends the upper limit of the integration to $\infty$ to obtain the following equations:
	\begin{align}
	\label{eq:4}
	&\dot{a}(t)=-\frac{i}{2}\sum^{}_{n}g^{}_{n}e^{i\left(E^{}_{0}t+\phi(t)-\varepsilon^{}_{n}t\right)}b^{}_{n}(t),\nonumber\\
	&\dot{b}^{}_{n}(t)=-\frac{i}{2}g^{}_{n}e^{-i\left(E^{}_{0}t+\phi(t)-\varepsilon^{}_{n}t\right)}a(t)-\frac{1}{4}\sum^{}_{l}C^{}_{nl}e^{i(\varepsilon^{}_{n}-\varepsilon^{}_{l})t}b^{}_{l}(t),
	\end{align}
	with $C^{}_{nl}=\int^{\infty}_{0}K^{}_{nl}(\tau)d\tau$. The non-diagonal terms ($n\neq l$) couple different TLSs. These could become important if two or more TLSs are close to degeneracy, $|\varepsilon^{}_{n}-\varepsilon^{}_{l}|\ll \gamma^{}_{l},\gamma^{}_{n}$. Then, cooperative super-radiant and sub-radiant dynamics of this group of TLSs could emerge~\cite{Dicke}. In our case this possibility can be neglected, as the average level spacing $\Delta$ between the TLSs is much larger than the typical width $\gamma^{}_{\mathrm{typ}}$. We therefore retain only diagonal terms, which gives   
	\begin{align}
	\label{eq:5}
	&\dot{a}(t)=-\frac{i}{2}\sum^{}_{n}g^{}_{n}e^{i\left(E^{}_{0}t+\phi(t)-\varepsilon^{}_{n}t\right)}b^{}_{n}(t),\nonumber\\
	&\dot{b}^{}_{n}(t)=-\frac{i}{2}g^{}_{n}e^{-i\left(E^{}_{0}t+\phi(t)-\varepsilon^{}_{n}t\right)}a(t)-\frac{1}{4}C^{}_{nn}b^{}_{n}(t).
	\end{align}
	Finally, we define $\tilde{b}^{}_{n}(t)=e^{C^{}_{nn}t/4}b^{}_{n}(t)$ and obtain
	\begin{align}
	\label{eq:6}
	&\dot{a}(t)=-\frac{i}{2}\sum^{}_{n}g^{}_{n}e^{i\left[E^{}_{0}t+\phi(t)-(\varepsilon^{}_{n}-iC^{}_{nn}/4)t\right]}\tilde{b}^{}_{n}(t),\nonumber\\
	&\dot{\tilde{b}}^{}_{n}(t)=-\frac{i}{2}g^{}_{n}e^{-i\left[E^{}_{0}t+\phi(t)-(\varepsilon^{}_{n}-iC^{}_{nn}/4)t\right]}a(t).
	\end{align}
	By absorbing the imaginary part of $C^{}_{nn}$ (Lamb shift) into the definition of $\varepsilon^{}_{n}$, and defining $\gamma^{}_{n}\equiv\operatorname{Re}(C^{}_{nn})/4$, one ends up with Eqs.~(4) of the main text [with $\tilde{b}^{}_{n}(t)$ replaced by $b^{}_{n}(t)$]. By integrating out the phonons, one thus finds an exact description of the system in terms of the non-Hermitian Hamiltonian~(2) of the main text.
	
	\subsection{Strong-coupling regime - additional figures} \label{strong}
\begin{figure}[ht!]
		\includegraphics[width=0.73\textwidth,height=0.34\textheight]{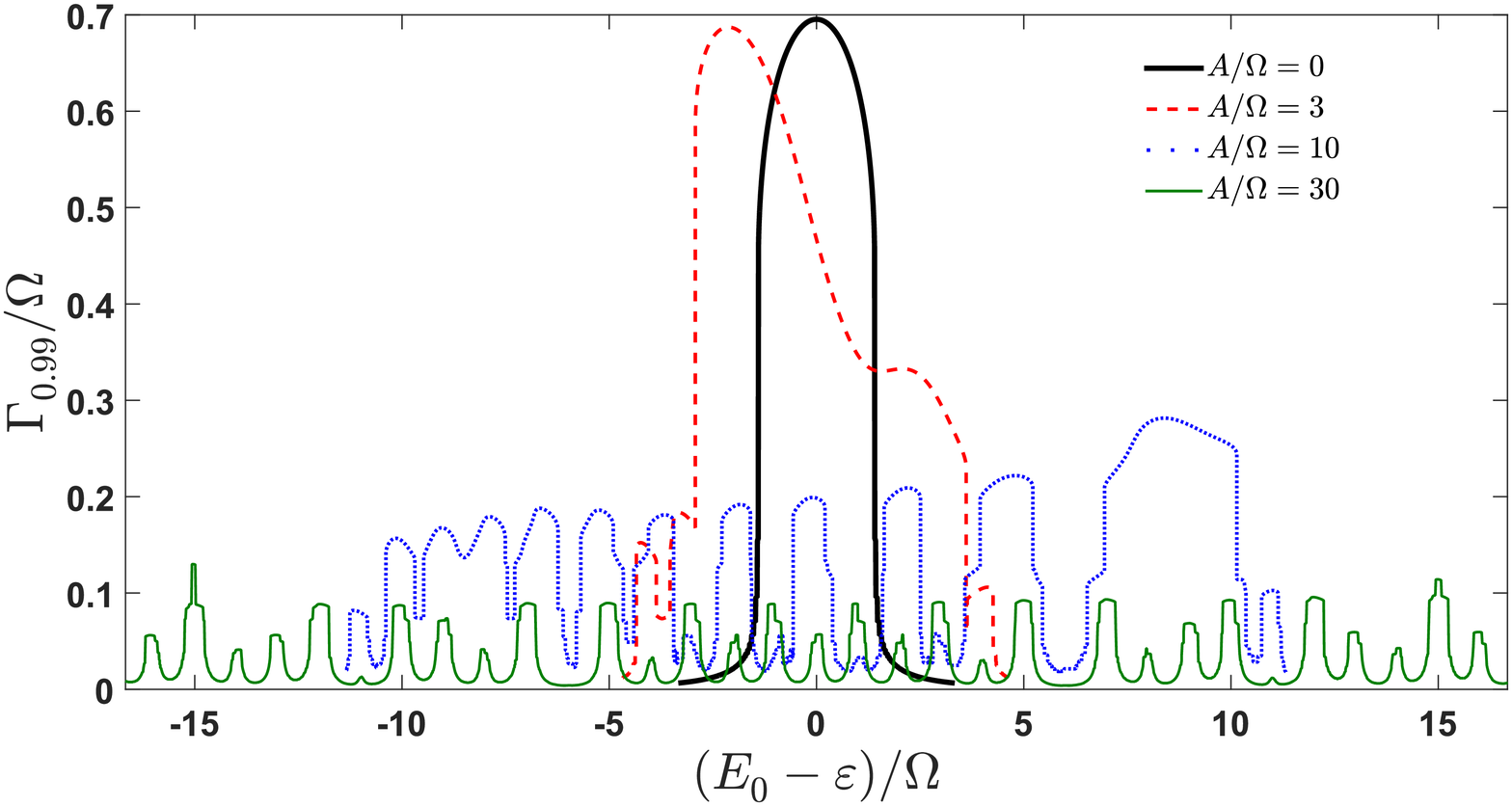}
		\caption{(Color online) Qubit relaxation rate as a function of its central energy splitting $E^{}_{0}$ (relative to the TLS energy splitting $\varepsilon$), obtained from a numerical integration of the Schr\"{o}dinger equation [Eqs.~(4) of the main text] for various values of $A/\Omega$, as indicated in the legend. The relaxation rate $\Gamma^{}_{0.99}$ (in units of $\Omega$) is calculated as $1/T$, where $T$ is the time for which $|a(T)|=0.99$. The calculation is performed for a single strongly coupled TLS, with $g/\Omega=0.2$ and $\gamma/\Omega=2/3\cdot 10^{-1}$.}
		\label{fig_supp1}
	\end{figure}

	In the main text, we have discussed the strong-coupling limit $g^{}_{n}\gg\gamma^{}_{n}$ assuming the modulation frequency $\Omega$ to be much larger then the typical coupling, meaning that terms with $m^{}_{1}\neq m^{}_{2}$ can be neglected in Eq.~(5) of the main text. This effectively transforms the contribution of a single strongly coupled TLS with energy splitting $\varepsilon^{}_{n}$ to a collection of TLSs at energy splittings $\tilde{\varepsilon}^{}_{nm}=\varepsilon^{}_{n}+m\Omega$ with smaller couplings $\tilde{g}^{}_{nm}=g^{}_{n}J^{}_{m}(A/\Omega)<g^{}_{n}$. This result is demonstrated in Fig.~1(b) of the main text for a single TLS, with $g/\Omega=2/3\cdot 10^{-1}$. If, however, the typical coupling is not much smaller than $\Omega$, the relaxation rate that follows from Eq.~(5) of the main text shows a more complex behavior as a function of the qubit central frequency $E^{}_{0}$. In this case, terms with $m^{}_{1}\neq m^{}_{2}$ cannot be neglected, which leads to multiple contributions for a given replication at $\tilde{\varepsilon}^{}_{nm}=\varepsilon^{}_{n}+m\Omega$. Such a case is shown in Fig.~\ref{fig_supp1}, which presents the same calculation as in Fig.~1(b) of the main text, but with $g/\Omega=0.2$ and $\gamma/\Omega=2/3\cdot 10^{-1}$ (instead of $g/\Omega=2/3\cdot 10^{-1}$ and $\gamma/\Omega=0.02$). Compared to Fig.~1(b) of the main text, one observes that the shape of the replicated peaks here is more complicated. In addition, the rate in Fig.~\ref{fig_supp1} is less symmetric with respect to the origin, because it is more sensitive to the phase of the frequency modulation. Since the qubit energy splitting is $E(t)=E^{}_{0}+A\cos(\Omega t)$, the qubit initially approaches resonance for $E^{}_{0}>\varepsilon$, whereas it moves away from resonance for $E^{}_{0}<\varepsilon$. Therefore, its relaxation rate is smaller for $E^{}_{0}=\varepsilon-\delta$ than for $E^{}_{0}=\varepsilon+\delta$, for any value of $\delta>0$. Note, however, that the main effect of the qubit frequency modulation remains also in Fig.~(\ref{fig_supp1}), i.e., the weight of the replications is reduced with increasing $A/\Omega$, and for $A/\Omega$ large enough each replication is Lorentzian as in the weak-coupling regime.
	
\begin{figure}[ht!]
		\includegraphics[width=0.73\textwidth,height=0.34\textheight]{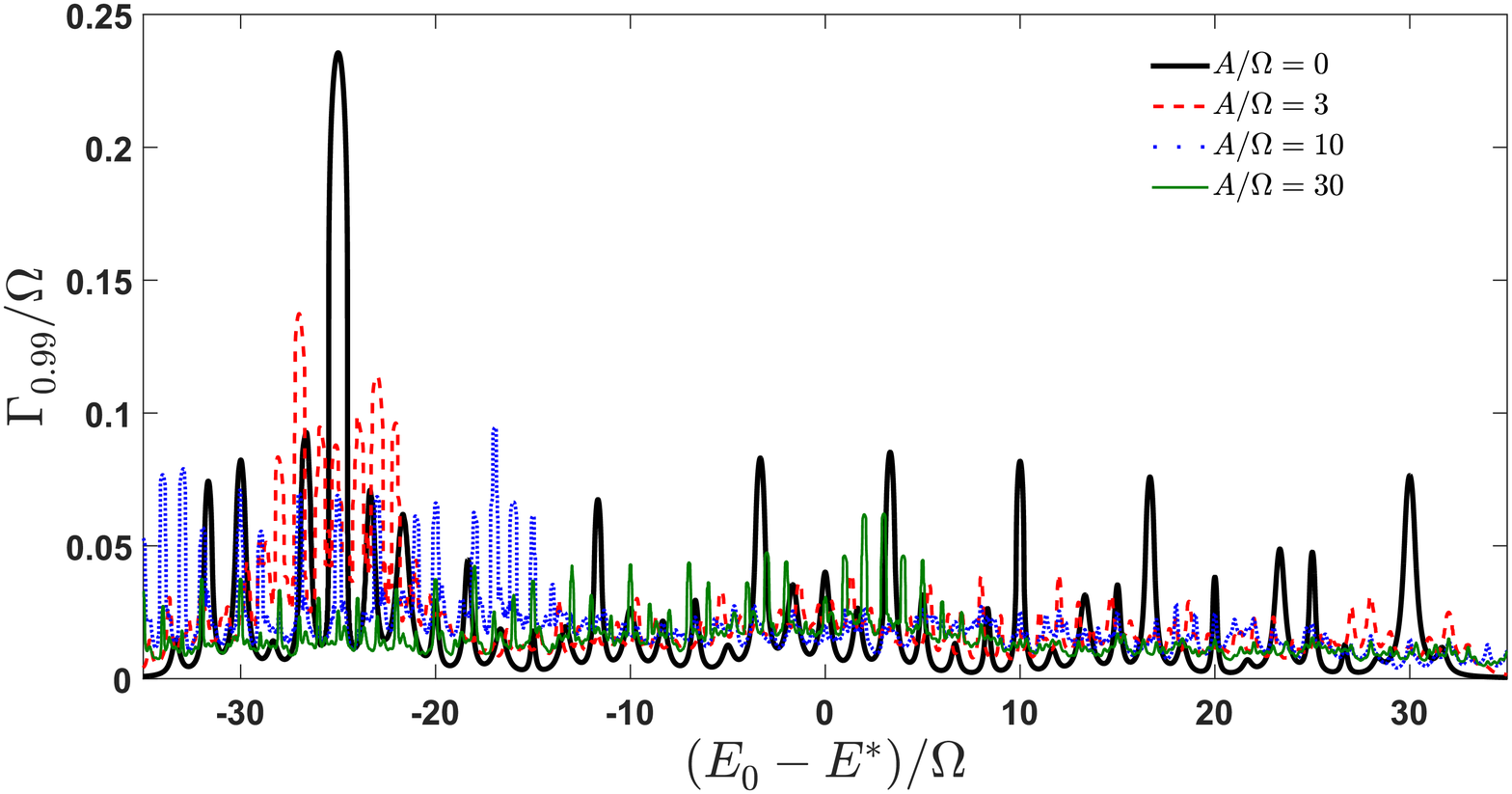}
		\caption{(Color online) Qubit relaxation rate as a function of its central energy splitting $E^{}_{0}$ (counted relative to some typical energy splitting of the qubit, $E^{\ast}$), obtained from a numerical integration of the Schr\"{o}dinger equation [Eqs.~(4) of the main text] for various values of $A/\Omega$, as indicated in the legend. The relaxation rate $\Gamma^{}_{0.99}$ (in units of $\Omega$) is calculated as $1/T$, where $T$ is the time for which $|a(T)|=0.99$. The calculation is performed for a bath of 41 TLSs with a constant level spacing $\Delta/\Omega=5/3$, 40 weakly coupled TLSs with couplings $g^{}_{n}/\Omega$ and relaxation rates $\gamma^{}_{n}/\Omega$ randomly chosen from homogeneous distributions in the ranges $[2/3,10/3]\cdot 10^{-2}$ and $[2/3,10/3]\cdot 10^{-1}$, respectively [similar realization as in Fig.~1(a) of the main text], and a single strongly coupled TLS at $(E^{}_{0}-E^{\ast})/\Omega=-25$ with $g/\Omega=2/3\cdot 10^{-1}$ and $\gamma/\Omega=0.02$ [as in Fig.~1(b) of the main text].}
		\label{fig_supp2}
	\end{figure}

	Another case that may be relevant to practical systems is a bath combining sparsely distributed weakly coupled TLSs and individual strongly coupled TLSs. In Fig.~(\ref{fig_supp2}) we show the resulting relaxation rate of a qubit coupled to 40 weakly coupled TLSs [with couplings and relaxation rates as in Fig.~1(a) of the main text] and a single strongly coupled TLS [with $g/\Omega=2/3\cdot 10^{-1}$ and $\gamma/\Omega=0.02$ as in Fig.~1(b) of the main text]. As expected, averaging of the fluctuations of the qubit relaxation rate is also observed in this case.
	
	\subsection{Qubit gates in the presence of frequency modulation} \label{Gates}
	In the main text we have argued that modulating the TLS frequencies has the advantage of allowing qubit gates as usual. On the other hand, if modulation of the qubit frequency is inevitable, it is necessary to adjust the pulses of the gates accordingly, as we now elaborate. Consider a single qubit gate performed by driving the qubit into Rabi oscillations. In the presence of a frequency modulation, the corresponding qubit Hamiltonian is
	\begin{align}
	\label{eq:7}
	&\mathcal{H}^{}_{\mathrm{Qubit}}(t)=-\frac{1}{2}\left[E^{}_{0}+A\cos(\Omega t)\right]\sigma^{}_{z}+\Omega^{}_{\mathrm{R}}\cos\left(\omega^{}_{\mathrm{d}}t\right)\sigma^{}_{x},
	\end{align}
	where $\omega^{}_{\mathrm{d}}$ is the driving frequency and $\Omega^{}_{\mathrm{R}}$ is the Rabi frequency. Treating $V(t)=\Omega^{}_{\mathrm{R}}\cos\left(\omega^{}_{\mathrm{d}}t\right)\sigma^{}_{x}$ as a perturbation, in the interaction picture it transforms into
	\begin{align}
	\label{eq:8}
	&V^{}_{\mathrm{I}}(t)=e^{-\frac{i}{2}\left(E^{}_{0}+\phi(t)\right)\sigma^{}_{z}}V(t)e^{\frac{i}{2}\left(E^{}_{0}+\phi(t)\right)\sigma^{}_{z}}=\Omega^{}_{\mathrm{R}}\cos\left(\omega^{}_{\mathrm{d}}t\right)\left[e^{-i\left(E^{}_{0}+\phi(t)\right)}\sigma^{}_{+}+e^{i\left(E^{}_{0}+\phi(t)\right)}\sigma^{}_{-}\right].
	\end{align}
	Using the Fourier series expansion $e^{i\phi(t)}=\sum^{\infty}_{m=-\infty}J^{}_{m}(A/\Omega)e^{im\Omega t}$, one observes that on resonance ($\omega^{}_{\mathrm{d}}=E^{}_{0}$) the Rabi frequency is multiplied by $J^{}_{0}(A/\Omega)$, such that the effective Rabi frequency is $\Omega^{}_{\mathrm{R},0}=\Omega^{}_{\mathrm{R}}J^{}_{0}(A/\Omega)$. Alternatively, if one drives the Rabi oscillations at frequency $\omega^{}_{\mathrm{d}}=E^{}_{0}+m\Omega$, then the the effective Rabi frequency is $\Omega^{}_{\mathrm{R},m}=\Omega^{}_{\mathrm{R}}J^{}_{m}(A/\Omega)$. Interestingly, if one drives in-phase at multiples harmonics of $\Omega$ with weights given by $J^{}_{m}(A/\Omega)$, i.e., if $V=\Omega^{}_{\mathrm{R}}\sum^{}_{m}J^{}_{m}(A/\Omega)\cos(\omega^{}_{\mathrm{d}}+m\Omega)\sigma^{}_{x}$ with $\omega^{}_{\mathrm{d}}=E^{}_{0}$, then under the rotating wave approximation we find $V^{}_{\mathrm{I}}=(1/2)\Omega^{}_{\mathrm{R}}\sigma^{}_{x}$, and the desired Rabi frequency $\Omega^{}_{\mathrm{R}}$ is achieved.


\begin{references}
\bibitem{Remark2} The fact that $\sum^{\infty}_{m=-\infty}J^{4}_{m}(A/\Omega)\sim\Omega/A$ results from the typical value $|J^{}_{m}(x)|\sim\sqrt{\Omega/A}$ of the Bessel functions for $|m|<x$.
\end{references}
\end{document}